\documentclass[pteplogo,hdvipdfmx]{ptephy_v1}%%%%%% to generate preprint number with ptep logo

\preprintnumber{XXXX-XXXX} %%% %%% Insert preprint number here

\usepackage{amssymb}
\usepackage{amsmath}
\usepackage[all, warning]{onlyamsmath}

\newcommand{\vev}[1]{\langle #1 \rangle}
\newcommand{\diracslash}[1]{#1 \hspace{-5pt}/ \:}
\newcommand{\Pslash}{P \hspace{-7pt}/ \:}

\usepackage[normalem]{ulem} 
\usepackage{color}

\font\arial=phvr8t at 10pt

\begin{document}

\title{Inverse mass ordering of light scalar mesons in the Nambu Jona-Lasinio model}

%%%% To generate auto affiliation numbers please use \author{}\affil{} command

\author[1]{Takahiro~Saionji}
\affil{Department of Physics, Tokyo Institute of Technology, Tokyo 152-8551, Japan}

\author[1]{Daisuke~Jido}
%\email{jido@th.phys.titech.ac.jp}

\author[2]{Masayasu~Harada} %%% Use optional bracket [3] to change the respective address
\affil{Department of Physics, Nagoya University, Nagoya, 464-8602, Japan}
\affil{Kobayashi-Maskawa Institute for the Origin of Particles and the Universe, Nagoya University, Nagoya, 464-8602, Japan}
\affil{Advanced Science Research Center, Japan Atomic Energy Agency, Tokai 319-1195, Japan}

%\author{Insert last author name here\thanks{These authors contributed equally to this work}}
%\affil{Insert last author address here}

%%% To include the collaborator name... Please use the command "\collaborator"
%%% For example: \collaborator{ATLAS Collaboration}

\begin{abstract}
The masses of the low-lying scalar mesons are investigated in the three-flavor Nambu Jona-Lasinio (NJL) model
by treating the scalar mesons as composite objects of a quark and an antiquark. It is known that 
a simple $\bar qq$ picture fails to reproduce so-called inverse mass ordering for the scalar mesons.
Recently a new mechanism to reproduce the observed mass spectrum of the scalar mesons was proposed 
in a linear sigma model by introducing flavor symmetry breaking induced by the U(1) axial anomaly. 
Motivated by this proposal, we examine whether this new mechanism works also 
in the NJL model. By calculating the scalar meson masses, we find that the NJL model 
reproduces the observed mass ordering with sufficient strength of the new term. 
With this mechanism, it turns out that the constituent strange quark mass gets degenerate 
to that of the up and down quark if the inverse mass ordering is reproduced. 
We also discuss the scalar diquark masses to check the consistency 
of the degeneracy of the constituent quark masses with the light baryon masses.
\end{abstract}

\subjectindex{
D32,%Hadron structure and interactions
B60 %Chiral lagrangians
}

%\begin{keyword}
%dynamical breaking of chiral symmetry \sep chiral effective models \sep
%scalar meson masses
%%U$_{A}$(1) anomaly \sep $\eta^{\prime}$ meson \sep scalar meson
%\end{keyword}

\maketitle

%% main text
\section{Introduction}
%\label{}

The understanding of the structure of the low-lying scalar mesons is important in hadron physics.
So far, several scalar mesons have been observed below 1 GeV: $f_{0}(500)$ with $I=0$, 
$K_{0}^{*}(700)$ with $I=1/2$, $f_{0}(980)$ with $I=0$ and $a_{0}(980)$ with $I=1$
as listed in the particle data table~\cite{PDG}. 
One of the characteristic feature of these scalar mesons is that, unlike the vector mesons, 
the mass of the isovector $a_{0}(980)$ meson is not close to that of the isoscalar 
$f_{0}(500)$ meson, and rather the $a_{0}(980)$ meson being non-strange
is heavier than the strange $K_{0}^{*}(700)$ meson.
It is expected that mesons are composed of a quark and an antiquark. Nevertheless, 
naive quark models based on the $\bar qq$ picture cannot explain 
the mass ordering of these light scalar mesons, because the masses of hadrons 
containing strange quarks should be heavier than those without strange quarks there. 
It has been suggested that 
this ``inverse mass ordering'' of the light scalar mesons can be explained by 
the tetraquark picture $\bar q\bar q qq$~\cite{Jaffe:1976ig}.
In this picture the isovector meson can have a hidden $\bar ss$ component and it gives a heavier mass to the meson.   
Apart from that, 
these scalar mesons have strong coupling to two pseudoscalar mesons and eventually 
have a large decay width. Such unstable resonances are described in hadron-hadron 
scattering. For instance, the $f_{0}(500)$ meson can be regarded as a $s$-wave resonance
in the isosinglet pion-pion scattering~\cite{Harada:1995dc,Dobado:1996ps,Oller:1997ng,Oller:1998hw,Igi:1998gn}. 
In these ways, the structure of the scalar mesons being controversial,
many pictures for the scalar mesons have been proposed so far such as 
two-quark $\bar qq$, four-quark $\bar q \bar q qq$, glueball, meson-meson scattering and their mixture~\cite{Black:1998wt,Ishida:1999qk,Kunihiro:2003yj,Chen:2006hyp,Zhang:2006xp,Pennington:2007yt,Klempt:2007cp,Kojo:2008hk,Fariborz:2009cq,Hyodo:2010jp,Parganlija:2012fy,Pelaez:2015qba,Pelaez:2016klv,Achasov:2017ozk,KHMJ}. 

Another aspect for the nature of scalar mesons is that there must exist the scalar fields transformed from the pion fields under chiral transformation in quantum chromodynamics (QCD),
that is chiral partners of the pions~\cite{GellMann:1960np}. 
Chiral symmetry is an approximated symmetry in QCD and is broken dynamically by
the flavor-singlet scalar condensate $\langle \bar qq \rangle$ with the emergence of 
the pseudoscalar Nambu-Goldstone bosons corresponding to the pion and its flavor partners.
The fluctuation mode of the quark condensate, the $\sigma$ meson, is regarded as 
one of the signals for the dynamical breaking of chiral symmetry. 
It is not understood yet how the sigma meson should appear in the observed spectrum. 
Here we have the reason that we adhere to the $\bar qq$ picture for the scalar mesons. 

So far there are several attempts to reproduce the inverse mass ordering in the $\bar qq$
picture~\cite{Volkov:1984kq,Klimt:1989pm,Vogl:1989ea,Ishida:1999qk,Dmitrasinovic:2000ei,Naito:2002,Osipov:2004bj,Su:2007au,Osipov:2012kk,Osipov:2013fka}. 
The key ingredient is the interaction induced by the U(1) axial anomaly~\cite{Kobayashi:1970ji,Kobayashi:1971qz,Schechter1971,'tHooft1976,tHooft:1976rip,Rosenzweig:1979ay,DiVecchia:1980yfw,Witten:1980sp,Kawarabayashi1980}, 
so-called Kobayashi-Maskawa-'t Hooft (KMT) interaction. 
For instance, Ref.~\cite{Naito:2002} employed an extended three-flavor Nambu Jona-Lasinio (NJL) model with the six-point KMT interaction to study the mass spectrum of the 
light scalar mesons. The KMT interaction works for the flavor singlet scalar meson
as attractive interaction, while for the octet scalar meson it behaves as repulsive 
interaction. This term, therefore, contributes to the isosinglet scalar meson to reduce its mass 
lighter than the other scalar mesons with $I=1$ and $I=1/2$. This is consistent with 
the mass ordering between $f_{0}(500)$ and the other scalar mesons. The flavor SU(3) 
breaking is introduced to the NJL model by the mass difference of the current quarks 
and enters to the meson mass spectrum through the constituent quark masses and 
the quark condensates. Owing to the flavor singlet nature of the KMT interaction, 
the KMT interaction works for the strange isodoublet meson 
with the up or down quark condensate,
while it contributes to the isovectore meson with the strange quark condensate. 
Because the absolute value of the strange quark condensate is larger than those of the
up and down condensates if the perturbative contributions are included, 
the repulsion of the KMT interaction is larger for the isovector 
meson than the isodoublet meson. This is consistent with the inverse mass ordering. 
Nevertheless, the flavor symmetry breaking in the quark condensates is not so sufficient
as to reproduce the inverse ordering of the masses for the $I=1$ and $I=1/2$ mesons.
Reference~\cite{Ishida:1999qk} was able to reproduce the inverse mass ordering of the 
light scalar mesons in a linear sigma model with the KMT interaction by introducing 
unsatisfactorily strong flavor symmetry breaking on the meson decay constants. 
In an NJL-like model~\cite{Osipov:2012kk,Osipov:2013fka}, by introducing a large number
of terms including explicit flavor symmetry breaking with the current quark mass, 
the inverse mass ordering was reproduced. Recently Ref.~\cite{KHMJ} proposed a new 
mechanism to realize the inverse mass ordering of the scalar mesons in a linear sigma model 
with a U(1) axial anomaly induced term including the current quark mass. 
The anomaly induced term with the current quark mass works similarly in the original KMT 
interaction for the scalar meson masses but with the current quark mass instead of the 
quark condensate. Thanks to a large flavor symmetry breaking in the current 
quark masses, say $m_{s}/m_{q} \sim 25$, the repulsive anomaly induced interaction 
contributes to the isovector meson much more than to the isodoublet meson.
Consequently the mass of $a_{0}(980)$ is provided larger than that of $K_{0}(700)$ in this model.

In this paper we utilize the Nambu Jona-Lasinio model~\cite{Nambu:1961tp}
to study the masses of the light scalar mesons in the $\bar qq $ picture.
Motivated by Ref.~\cite{KHMJ} for the linear sigma model, we introduce the chiral anomaly
term with the current quark mass to calculate the meson masses. 
We examine whether this term works to reproduce the
inverse mass ordering for the light scalar mesons also in the NJL model. 

The paper is organized as follows: In Sect.~\ref{sec:model}, the formulation of 
this work is explained. 
In Sect.~\ref{sec:results}, we show our numerical results. In Sect.~\ref{sec:discussion} we discuss the diquark masses and 
the Gell-Mann Oakes Renner relation in this model. 
Finally Sect.~\ref{sec:summary} is devoted for the conclusion of this paper. In Appendix calculation details 
are shown.

%%%
\section{Nambu Jona-Lasinio model}  \label{sec:model}
In order to investigate the masses of the scalar mesons, 
we consider the Nambu Jona-Lasinio model (NJL model) 
for the three flavors~\cite{Bernard:1987gw,Kunihiro:1987bb,Vogl:1991qt,Hatsuda:1994pi}. 
The Lagrangian that we use here for the quark field $\psi = (u,d,s)^{T}$ is 
given as
\begin{align}
  {\cal L}_\textrm{NJL} =& \
   \bar \psi ( i \diracslash{\partial} -{\cal M}) \psi
    + g_{S} \sum_{A=0}^{8}\left[ \left(\bar \psi \frac{\lambda_{A}}2 \psi\right)^{2} 
    + \left(\bar \psi i \gamma_{5} \frac{\lambda_{A}}2 \psi\right)^{2} \right] 
    \nonumber \\ & \ 
%    + \frac{g_{D}}2 \left\{ \epsilon_{ijk} \epsilon_{abc} 
%    \bar \psi_{i} (1+\gamma_{5}) \psi_{\beta} \bar \psi_{a} (1+\gamma_{5}) \psi_{b} \bar \psi_{k} (1+\gamma_{5}) \psi_{c} + (\gamma_{5} \rightarrow - \gamma_{5}) \right\}
    + \frac{g_{D}}2 \left\{ \det[ \bar \psi_{i} (1+\gamma_{5}) \psi_{j}]
    + \det[ \bar \psi_{i} (1-\gamma_{5}) \psi_{j}] \right\}
    \nonumber \\ & \ 
    + \frac{g_{k}}4 \left\{ \epsilon_{ijk} \epsilon_{abc} 
    {\cal M}_{ia} \bar \psi_{j} (1+\gamma_{5}) \psi_{b} \bar \psi_{k} (1+\gamma_{5}) \psi_{c} + (\gamma_{5} \rightarrow - \gamma_{5}) \right\}, \label{NJLLag}
\end{align}
%\begin{equation}
%\begin{split}
%  {\cal L}_\textrm{ NJL} =& \
%   \bar \psi ( i \diracslash{\partial} -{\cal M}) \psi
%    + g_{S} \sum_{A=0}^{8}\left[ \left(\bar \psi \frac{\lambda_{A}}2 \psi\right)^{2} 
%    + \left(\bar \psi i \gamma_{5} \frac{\lambda_{A}}2 \psi\right)^{2} \right] 
%     \\ & \ 
%    + \frac{g_{D}}2 \left\{ \det[ \bar \psi_{i} (1+\gamma_{5}) \psi_{j}]
%    + \det[ \bar \psi_{i} (1-\gamma_{5}) \psi_{j}] \right\}
%    \\ & \ 
%    + \frac{g_{k}}4 \left\{ \epsilon_{ijk} \epsilon_{abc} 
%    {\cal M}_{ia} \bar \psi_{j} (1+\gamma_{5}) \psi_{b} \bar \psi_{k} (1+\gamma_{5}) \psi_{c} + (\gamma_{5} \rightarrow - \gamma_{5}) \right\} \label{NJLLag}
%\end{split}
%\end{equation}
with the quark mass matrix ${\cal M} = \textrm{ diag}(m_{q}, m_{q}, m_{s})$, 
the model parameters $g_{S}$, $g_{D}$, $g_{k}$,
the Gell-Mann matrices $\lambda^{A}$ ($A=0,1, \dots, 8$) normalized as 
$\textrm{Tr}[\lambda_{A} \lambda_{B}] = 2 \delta^{AB}$, and 
the antisymmetric tensor $\epsilon_{ijk}$ for $i,j,k = 1,2,3$. 
In this Lagrangian, it is understood that the determinant is taken in the flavor space, 
the latin letters are the flavor indices of the quark fields running from 1 to 3,
and the summations is taken over the repeated indices. 
We assume isospin symmetry by $m_{q} = m_{u} = m_{d}$.
We call the up and down quark fields collectively
by $q$ and the strange quark field by $s$. 

The last two terms in Lagrangian~\eqref{NJLLag} are originated from the chiral anomaly in QCD.
The term with $g_{D}$, known as the Kobayashi Maskawa t'Hooft interaction, 
provides six-point vertices, and is invariant 
under the SU(3)$_{L} \otimes$SU(3)$_{R}$ but breaks the U(1)$_{A}$ symmetry. 
With the presence of this term the $\eta^{\prime}$ meson is degraded from a Nambu Goldstone 
boson associated with spontaneous breaking of chiral symmetry. 
The term containing $g_{k}$ is also an anomalous interaction and provides four-point interactions
of quarks. It breaks chiral symmetry explicitly with the quark mass term $\cal M$ and 
the flavor symmetry is also broken 
by the quark mass difference $m_{q} \neq m_{s}$. 
This term is uniquely determined by the U(1)$_{A}$ anomaly 
and the SU(3) flavor symmetry with one quark mass term $\cal M$. 
This type of interactions was introduced 
into a linear sigma model in Ref.~\cite{KHMJ} to investigate the inverse mass hierarchy 
of the scalar mesons and is also found in Ref.~\cite{Osipov:2012kk} as $L_{2}$
for the NJL-like model. 
Because containing the totally antisymmetric tensor $\epsilon_{ijk}$ in the flavor space,
this term contributes to non-strange systems with the strange quark mass
and to systems having the strange quarks with the light quark mass. 
Therefore, the term provides significantly stronger interactions for the non-strange systems. 

Employing the mean field approximation~\cite{Hatsuda:1994pi}
by replacing the quark fields in the Lagrangian as
\begin{align}
%    \bar \psi_{1}  \psi_{1} \bar \psi_{2} \psi_{2}  \to\ &  
%   \langle \bar \psi_{1}  \psi_{1} \rangle  \bar \psi_{2} \psi_{2}
%   + \langle \bar \psi_{2}  \psi_{2} \rangle  \bar \psi_{1} \psi_{1} 
%   - \langle \bar \psi_{1}  \psi_{1} \rangle  \langle \bar \psi_{2} \psi_{2}\rangle, \\
%   \bar \psi_{1}  \psi_{1} \bar \psi_{2} \psi_{2} \bar \psi_{3}\psi_{3}  \to\ &
%   \langle \bar \psi_{1}  \psi_{1} \rangle \langle \bar \psi_{2} \psi_{2} \rangle \bar \psi_{3} \psi_{3}
%   + \langle \bar \psi_{3}  \psi_{3} \rangle \langle \bar \psi_{1} \psi_{1}\rangle \bar \psi_{2} \psi_{2}
%   + \langle \bar \psi_{2}  \psi_{2} \rangle  \langle\bar \psi_{3} \psi_{3}\rangle \bar \psi_{1} \psi_{1} \nonumber \\
%   &- 2\langle \bar \psi_{1}  \psi_{1} \rangle  \langle \bar \psi_{2} \psi_{2}\rangle \langle \bar \psi_{3} \psi_{3} \rangle,
    \bar \psi_{i}  \psi_{i} \bar \psi_{j} \psi_{j}  \to &  \ 
   \langle \bar \psi_{i}  \psi_{i} \rangle  \bar \psi_{j} \psi_{j}
   + \langle \bar \psi_{j}  \psi_{j} \rangle  \bar \psi_{i} \psi_{i} 
   - \langle \bar \psi_{i}  \psi_{i} \rangle  \langle \bar \psi_{j} \psi_{j}\rangle, \\
   \bar \psi_{i}  \psi_{i} \bar \psi_{j} \psi_{j} \bar \psi_{k}\psi_{k}  \to &\ 
   \langle \bar \psi_{i}  \psi_{i} \rangle \langle \bar \psi_{j} \psi_{j} \rangle \bar \psi_{k} \psi_{k}
   + \langle \bar \psi_{k}  \psi_{k} \rangle \langle \bar \psi_{i} \psi_{i}\rangle \bar \psi_{j} \psi_{j}
   + \langle \bar \psi_{j}  \psi_{j} \rangle  \langle\bar \psi_{k} \psi_{k}\rangle \bar \psi_{i} \psi_{i}
   \nonumber \\ &\ 
   - 2\langle \bar \psi_{i}  \psi_{i} \rangle  \langle \bar \psi_{j} \psi_{j}\rangle \langle \bar \psi_{k} \psi_{k} \rangle,
\end{align}
where $\psi_{i}$ $(i=1,2,3)$ is either $u$, $d$ or $s$ quark field and the summation is not taken for the repeated indices,
we obtain the gap equations for the dynamical masses of the light and strange quarks, 
$M_{q}$ and $M_{s}$, as
\begin{subequations} \label{gapeq}
\begin{align}
   M_{q} &= m_{q} - g_{S} \vev{\bar qq} - g_{D} \vev{\bar qq}\vev{\bar ss}
   - g_{k} (m_{q} \vev{\bar ss} + m_{s} \vev{qq}), \label{gapq} \\
   M_{s} &= m_{s} - g_{S} \vev{\bar ss} - g_{D} \vev{\bar qq}^{2}
   -2g_{k} m_{q} \vev{\bar qq},  \label{gaps}
\end{align}
\end{subequations}
with the isospin symmetry $\vev {\bar qq} = \vev{\bar uu} = \vev{\bar dd}$. 

We evaluate the quark condensate from the quark propagator
$S_{F}(x) = -i \vev {0| T [ \psi(x) \bar \psi(x)] | 0}$
with the dynamical mass~$M$
as
\begin{equation}
   \vev{\bar \psi \psi} = - i N_{c} {\lim_{x \to 0}} \textrm{Tr} [S_{F}(x)]
      = - i N_c \int \frac{d^{4}q}{(2\pi)^{4}} \textrm{ Tr}\left[ \frac{1}{\diracslash{p} - M+i\epsilon } \right] , \label{qq}
\end{equation}
where $N_{c}$ is the number of color, that is $N_{c}=3$, 
and $\textrm{Tr}$ is taken for the Dirac indices. To regularize the momentum 
integral, we make use of the three-momentum cutoff $\Lambda$ in accordance with 
the suggestion given in Ref.~\cite{Hatsuda:1994pi}. 
The calculated result of Eq.~\eqref{qq} 
% with the cutoff~$\Lambda$
is given in Eq.~\eqref{eq:SF}. 
In the present work, the cutoff $\Lambda$, which determines 
the mass scale of the scalar meson, is to be fixed at $\Lambda = 650$~MeV commonly for three quarks. 
The qualitative consequences do not change in a small change of the value of the cutoff and the corresponding 
adjustment of the model parameters. 
With sufficiently large couplings in magnitude, the gap equations~\eqref{gapeq} provide nontrivial solutions where
chiral symmetry is dynamically broken. We introduce dimensionless coupling constants as
\begin{equation} \label{DLparam}
   G_{S}  = \frac{g_{S}}{g_{S}^{0}}, \qquad
   G_{D}  = \frac{\Lambda g_{D}}{(g_{S}^{0})^{2}} ,\qquad
   G_{k} = \frac{\Lambda g_{k}}{g_{S}^{0}} ,
%   \qquad \textrm{with} \quad g_{S}^0 = 2\pi^{2}/(3\Lambda^{2})
%   G_{S}  = \frac{g_{S}}{g_{S}^{0}} =  \frac{3\Lambda^{2}}{2\pi^{2}} g_{S}, \qquad
%   G_{D}  = \frac{\Lambda g_{D}}{(g_{S}^{0})^{2}} = \Lambda \left(\frac{3\Lambda^{2}}{2\pi^{2}}\right)^{2}g_{D},\qquad
%   G_{k} = \frac{\Lambda g_{k}}{g_{S}^{0}} = \Lambda \frac{3\Lambda^{2}}{2\pi^{2}} g_{k},
\end{equation}
with 
\begin{equation}
   g_{S}^0 = \frac{2\pi^{2}}{3\Lambda^{2}},
\end{equation}
%$$g_{S}^0 = 2\pi^{2}/(3\Lambda^{2}),$$ 
which is the critical coupling of the dynamical symmetry breaking 
for $g_{D}=g_{k}=0$ in the chiral limit. 

In order to calculate the meson masses, let us consider the quark-antiquark scattering amplitude $T$,
which is defined as
\begin{equation}
   T(P) = \int {d^{4}x} \langle 0 | T \Phi(x) \Phi(0) | 0 \rangle e^{-iP\cdot x}, \label{Tmatrix}
\end{equation}
where $\Phi(x)$ is an interpolating field of the meson of interest and is 
given by $\Phi(x) = \bar \psi_{i} \psi_{j}$ for the scalar meson and 
$\Phi(x) = \bar \psi_{i} i \gamma_{5} \psi_{j}$ for the pseudoscalar meson. 
The quark indices $i,j$ stand for the flavor of the quark and are appropriately chosen to provide 
the meson flavor.
For the isospin $I=0$ configurations of the flavor singlet and octet, 
we consider their quark contents to be $\Phi_{0}=\frac1{\sqrt 3} (\bar uu + \bar dd + \bar ss)$ and 
$\Phi_{8}=\frac 1{\sqrt 6}(\bar uu + \bar dd - 2\bar ss)$, respectively.
The explicit quark contents of the interpolating fields are summarized in Table~\ref{tab:QC}.
Here we call the isodoublet scalar meson $\kappa$ and the isovector scalar meson $a_{0}$. 

\begin{table}
\centering
\begin{tabular}{ll}
\hline\hline
\multicolumn{2}{c}{Scalar mesons}\\
\hline
$\Phi_{a_{0}} = \bar q \frac{\tau^{a}}{\sqrt 2} q$, & 
$\Phi_{\kappa} = \bar s q $, \\
$\Phi^{S}_{8} = \frac 1{\sqrt 6}(\bar uu + \bar dd - 2\bar ss)$, &
$\Phi^{S}_{0} = \frac1{\sqrt 3} (\bar uu + \bar dd + \bar ss)$,  \\
\hline
\multicolumn{2}{c}{Pseudoscalar mesons}\\
\hline
$\Phi_{\pi} = \bar q  i\gamma_{5}\frac{\tau^{a}}{\sqrt 2} q$, & 
$\Phi_{K} = \bar s  i\gamma_{5}q $, \\
$\Phi^{P}_{8} = \frac 1{\sqrt 6}(\bar u i\gamma_{5}u + \bar d i\gamma_{5}d - 2\bar s i\gamma_{5}s)$, &
$\Phi^{P}_{0} = \frac1{\sqrt 3} (\bar u i\gamma_{5}u + \bar d i\gamma_{5}d + \bar s i\gamma_{5}s)$,  \\
\hline\hline
\end{tabular}
\caption{Quark contents of the meson interpolating fields. 
$q$ is the isodoublet quark field and $\tau^{a}$ is the Pauli matrix for the isospin space. }  \label{tab:QC} 
\end{table}

The mesons are dynamically generated in quark-antiquark
scattering with sufficiently strong attraction  
between the quark and the antiquark, and are expressed as poles of the scattering amplitude~$T$.
We calculate the $T$-matrix by solving the Bethe-Salpeter equation:
\begin{equation}
   T = K +  K J T, \label{BSeq}
\end{equation}
with the interaction kernel $K$ and the quark loop function $J$.
Here we use the ladder approximation in the center of mass frame.

The interaction kernel $K$ is obtained directly from the four-point interactions 
of the $g_{S}$ and $g_{k}$ terms in Lagrangian~\eqref{NJLLag} and 
also from the six-point interaction with the mean field approximation. 
%The explicit form for each channel is shown in Table~\ref{intK}.
The explicit form of each channel is given as
\begin{subequations} \label{KScalar}
\begin{align}
K_{a_{0}} &= g_{S} - g_{D}\vev{\bar ss} - g_{k} m_{s}, \label{eq:a0K} \\ 
K_{\kappa} &= g_{S}-g_{D}\vev{\bar qq} - g_{k}m_{q},\\
K^{S}_{00} &= g_{S} + \frac 23 g_{D}(2\vev{\bar qq} + \vev{\bar ss}) + \frac 23 g_{k}(2m_{q} + m_{s}),\\
K^{S}_{88} &= g_{S} - \frac 13g_{D} (4\vev{\bar qq} - \vev{\bar ss}) - \frac13 g_{k}(4m_{q} - m_{s}),\\
K^{S}_{08} &= -\frac {\sqrt 2}3g_{D} (\vev{\bar qq} - \vev{\bar ss}) - \frac{\sqrt 2}{3}g_{k}(m_{q}-m_{s}), 
\end{align}
\end{subequations}
for the scalar mesons and
\begin{subequations} \label{KPScalar}
\begin{align}
K_{\pi} &= g_{S} + g_{D}\vev{\bar ss} + g_{k} m_{s},  \label{eq:pionK} \\ 
K_{K} &= g_{S}+g_{D}\vev{\bar qq} + g_{k} m_{q},\label{eq:kaonK} \\
K^{P}_{00} &= g_{S} - \frac 23 g_{D}(2\vev{\bar qq} + \vev{\bar ss}) - \frac 23 g_{k} (2m_{q} + m_{s}),\\
K^{P}_{88} &= g_{S} + \frac 13g_{D} (4\vev{\bar qq} - \vev{\bar ss}) + \frac13 g_{k} (4m_{q} - m_{s}),\\
K^{P}_{08} &= \frac {\sqrt 2}3 g_{D}(\vev{\bar qq} - \vev{\bar ss}) + \frac{\sqrt 2}{3} g_{k} (m_{q}-m_{s}),
\end{align}
\end{subequations}
for the pseudoscalar mesons. 
Thanks to the flavor symmetry breaking, $m_{q} \neq m_{s}$ and 
$\vev{\bar qq} \neq \vev{\bar ss}$,
we have the mixing between the flavor singlet and octet for $I=0$.

The loop function $J$ is defined with $P^{\mu}=(W, 0,0,0)$ by
\begin{equation}
   J_\textrm{S}(W;M_{1},M_{2}) = N_{c} i \int \frac{d^{4}p}{(2\pi)^{4}} \textrm{Tr} 
   \left[\frac{1}{\diracslash{p} - M_{1}+i\epsilon}\frac{1}{\diracslash{p} - \Pslash - M_{2}+i\epsilon}\right],
\end{equation}
for the scalar channel and
\begin{equation}
    J_{\textrm{P}}(W;M_{1},M_{2}) 
  =N_{c} i \int \frac{d^{4}p}{(2\pi)^{4}} 
  \textrm{Tr}\left[i\gamma_{5}\frac{1}{\diracslash{p} - M_{1}+i\epsilon} i\gamma_{5} \frac{1}{\diracslash{p} - \Pslash - M_{2}+i\epsilon}\right],
\end{equation}
for the pseudoscalar channel. We calculate
the loop function with the same three momentum cutoff $\Lambda$ as the calculation of the quark condensate.
This procedure keeps chiral symmetry in the calculation.
The explicit forms of the loop functions are shown in Eqs.~\eqref{eq:JS} for the scalar channel and 
\eqref{eq:JP} for the pseudoscalar channel. The loop function for each meson channel is listed as
\begin{subequations}  \label{loopJ}
\begin{align}
J_{a_{0}/\pi}(W) &= J_\textrm{S/P}(W; M_{q}, M_{q}),\\  
J_{\kappa/K}(W)  &= J_\textrm{S/P}(W; M_{q}, M_{s}),\\
J^{S/P}_{00}(W) &=  \frac 13 (2J_\textrm{S/P}(W; M_{q}, M_{q}) + J_\textrm{S/P}(W; M_{s}, M_{s})), \\
J^{S/P}_{88}(W) &=  \frac 13 (J_\textrm{S/P}(W; M_{q}, M_{q}) + 2J_\textrm{S/P}(W; M_{s}, M_{s})),\\
J^{S/P}_{08}(W) &=  \frac{\sqrt 2}3 (J_\textrm{S/P}(W; M_{q}, M_{q}) -  J_\textrm{S/P}(W; M_{s}, M_{s})).
\end{align}
\end{subequations}
%\begin{subequations}  \label{JScalar}
%\begin{align}
%J_{a_{0}}(W) &= J_\textrm{S}(W; M_{q}, M_{q}),\\  
%J_{\kappa}(W)  &= J_\textrm{S}(W; M_{q}, M_{s}),\\
%J^{S}_{0}(W) &=  \frac 13 (2J_\textrm{S}(W; M_{q}, M_{q}) + J_\textrm{S}(W; M_{s}, M_{s})), \\
%J^{S}_{8}(W) &=  \frac 13 (J_\textrm{S}(W; M_{q}, M_{q}) + 2J_\textrm{S}(W; M_{s}, M_{s})),\\
%J^{S}_{08}(W) &=  \frac{\sqrt 2}3 (J_\textrm{S}(W; M_{q}, M_{q}) -  J_\textrm{S}(W; M_{s}, M_{s})), 
%\end{align}
%\end{subequations}
%for the scalar channel and
%\begin{subequations}  \label{JPScalar}
%\begin{align}
%J_{\pi}(W) &= J_\textrm{P}(W; M_{q}, M_{q}) ,\\
%J_{K}(W)  &= J_\textrm{P}(W; M_{q}, M_{s}),\\
%J^{P}_{00}(W) &=  \frac 13 (2J_\textrm{P}(W; M_{q}, M_{q}) + J_\textrm{P}(W; M_{s}, M_{s})), \\
%J^{P}_{88}(W) &=  \frac 13 (J_\textrm{P}(W; M_{q}, M_{q}) + 2J_\textrm{P}(W; M_{s}, M_{s})), \\
%J^{P}_{08}(W) &=  \frac{\sqrt 2}3 (J_\textrm{P}(W; M_{q}, M_{q}) -  J_\textrm{P}(W; M_{s}, M_{s})), 
%\end{align}
%\end{subequations}
%for the pseudoscalar channel. 

The meson mass is obtained as the pole position of the scattering amplitude $T(W)$. 
In the ladder approximation the Bethe-Salpeter equation~\eqref{BSeq} can be solved in an 
algebraic way as 
\begin{equation}
T(W) = \frac{1}{1 - KJ(W)}K.
\end{equation}
We find the pole position of the scattering amplitude $T$ by solving
\begin{equation}
   1 - KJ(W) = 0, \label{eq:mass}
\end{equation}
for the single channel case. The loop function $J(W)$ in the complex $W$ plane has 
two Riemann sheets corresponding to the sign of the momentum
and the branch cut runs along the real axis from the threshold of the particle production 
$W=M_{1} + M_{2}$ to $+ \infty$. 
The solutions of Eq.~\eqref{eq:mass} found in the first Riemann sheet 
and below the threshold correspond to the bound states. 
The bound state solutions appear on the real axis.
On the other hand, solutions found 
in the second Riemann sheet correspond to virtual or 
resonance states; 
the virtual states are found below the threshold, while the resonance states
are above the threshold. 
Because of the lack of the confinement mechanism in the NJL model 
the mesons that are generated above the threshold have a decay to
fall apart into the quark and the antiquark. These decays are indeed model artifacts
of the NJL model.

With the flavor symmetry breaking, the $I=0$ channels of the flavor singlet and octet
couple each other and the Bethe-Salpeter equation becomes a matrix equation:
\begin{equation}
   \begin{pmatrix} T_{00} & T_{08} \\ T_{08} & T_{88}  \end{pmatrix}
   = \begin{pmatrix} K_{00} & K_{08} \\ K_{08} & K_{88}  \end{pmatrix}
   +  \begin{pmatrix} K_{00} & K_{08} \\ K_{08} & K_{88}  \end{pmatrix}
       \begin{pmatrix} J_{00} & J_{08}\\ J_{08} & J_{88}  \end{pmatrix}
       \begin{pmatrix} T_{00} & T_{08} \\ T_{08} & T_{88}  \end{pmatrix}. \label{eq:CCBS}
\end{equation}
The pole positions of the $T$-matrix are obtained for the coupled-channels case by
\begin{equation}
   \textrm{det}(1 - KJ(W)) = 0.
\end{equation}
We call the energetically lower solution for the scalar (pseudoscalar) meson by $\sigma$ $(\eta)$,
while the higher solution denotes $\sigma^{\prime}$ $(\eta^{\prime})$.

The decay constant $F$ for the pseudoscalar meson $\Phi^{a}$ is defined by 
\begin{align}
\langle 0 | A_{\mu}^{a}(x) | \Phi^{b} (P)\rangle &= i \delta^{ab} P_{\mu} F e^{-iP\cdot x}, \label{eq:decayconst} 
\end{align}
with
the axial vector current $A_{\mu}^{a} = \bar \psi \gamma_{\mu} \gamma_{5} \frac{\lambda^{a}}{2} \psi$, 
and calculated in the current NJL model as
\begin{equation}
   F = -i \sqrt{2R} J_{A}(m),
\end{equation}
with the meson mass $m$ and the residue $R$ of the $T$ matrix~\eqref{Tmatrix} as a function of
$W^{2}$ at $W^{2}=m^{2}$.
The loop function $J_{A}$ with the axial current is defined by 
\begin{equation}
   P^{\mu}J_{A}(W;M_{1},M_{2})  \equiv i N_{c} \int \frac{d^{4}p}{(2\pi)^{4}} 
   \textrm{Tr}\left[\frac{\gamma^{\mu}\gamma_{5}}2 \frac{1}{\diracslash{p}-M_{1} + i \epsilon} 
   i \gamma_{5} \frac{1}{\diracslash{p} - \Pslash - M_{2}+i \epsilon}\right].
\end{equation}
The loop function $J_{A}$ is also calculated with the same three-momentum cutoff $\Lambda$
as Eq.~\eqref{qq}. The explicit form is shown in Eq.~\eqref{eq:JA} in appendix.

\begin{table}
\centering
\begin{tabular}{cccc}
\hline\hline
$m_{\pi}$ & $m_{K}$ & $m_{\eta^{\prime}}$ & $f_{\pi}$ \\
\hline
$138.04$ & $495.65$ & $957.78$ & $92.2$ \\
\hline\hline
\end{tabular}
\caption{Input parameters in units of MeV.} \label{tab:inputs}
\end{table}

\section{Results}  \label{sec:results}
In this section, we show our calculated results. 
In Sect.~\ref{sec:param} we discuss the determined parameters. 
In Sect.~\ref{sec:scalar} we show our numerical results for the scalar meson masses.
In Sect.~\ref{sec:quark} we discuss the constituent quark masses for finite $G_{k}$. 

\begin{table}
\centering
\begin{tabular}{cl|rrrrr|c} 
\hline\hline
$G_{k}$ & & $0.0$ & $0.5$ & $1.0$ & $1.5$ & $2.0$ & PDG \cite{PDG} \\
\hline
$m_{q}$ & [MeV] & $  5.32$ & $  4.35$ & $  3.74$ & $  3.31$ & $  2.99$  & $3.45^{+0.35}_{-0.15}$ \\
$m_{s}$ & [MeV] & $134.79$ & $115.73$ & $101.39$ & $ 90.24$ & $ 81.34$  & $93.4^{+8.6}_{-3.4}$\\
$G_{S}$ & & $ 0.870$ & $ 0.845$ & $ 0.826$ & $ 0.812$ & $ 0.800$ & ---\\
$G_{D}$ & & $-0.865$ & $-0.768$ & $-0.692$ & $-0.630$ & $-0.579$ & ---\\
\hline
$m_{s}/m_{q}$ & & $ 25.35$ & $ 26.60$ & $ 27.15$ & $ 27.28$ & $ 27.17$ & $27.33^{+0.67}_{-0.77}$\\
\hline\hline
\end{tabular}
\caption{Determined parameters for $G_{k} = 0.0$, $0.5$, $1.0$, $1.5$ and $2.0$. 
The last column shows the values for current quark masses picked up from 
the Review of Particle Physics  (2022)~\cite{PDG}. 
The dimensionless interaction parameters, $G_{S}$, $G_{D}$ and $G_{k}$, are 
defined in Eq.~\eqref{DLparam}. The cutoff is fixed as $\Lambda = 650$~MeV.
} \label{tab:param}
\end{table}

\subsection{Determined parameters} \label{sec:param}
The model parameters, $m_{q}$, $m_{s}$, $G_{S}$ and $G_{D}$, 
are determined for each $G_{k}$ 
so as to reproduce the masses of $\pi$, $K$ and $\eta^{\prime}$ and the pion 
decay constant. The explicit values of these input quantities are shown in Table~\ref{tab:inputs}. 
Since the $\eta^{\prime}$ meson in this model is found as a resonance with a large width
above the threshold of the strange quark-antiquark production,
we reproduce the observed $\eta^{\prime}$ mass by the real part of the calculated resonance mass. 

The determined model parameters for $G_{k}=0.0$, $0.5$, $1.0$, $1.5$ and $2.0$ are shown 
in Table~\ref{tab:param}. 
We find that the values of the current quark masses $m_{q}$ and $m_{s}$ and their ratio for $G_{k}=2.0$ 
are more consistent with the values given in the Review of Particle Physics (2022)~\cite{PDG}
than those for $G_{k} =0.0$.
In Table~\ref{tab:param} we also find that the sizes of all the model parameters get smaller for larger $G_{k}$.
In particular, the current strange quark mass $m_{s}$ and the magnitude of $G_{D}$ are more strongly reduced.
These behaviors can be explained by the following argument.
We determine the model parameters so as to reproduce the observed masses of pion and kaon. 
The meson masses are obtained as the pole position of the $T$-matrix, which is 
calculated by the Bethe-Salpeter equation~\eqref{BSeq} with the interaction kernel $K$.
To keep the pion mass as the $G_{k}$ increases, the interaction kernel for pion should 
also stay as it is for $G_{k}=0$. According to Eq.~\eqref{eq:pionK}, 
as the effect of the $G_{k}$ term increases with a massive $m_{s}$,
the effects of the $G_{S}$ and $G_{D}$ terms should be reduced, and thus, 
the sizes of $G_{S}$ and $G_{D}$ have to decrease.
Note in Eq.~\eqref{eq:pionK} that both $G_{D}$ and $\vev{\bar ss}$ have a negative value. 
Because the $G_{k}$ term comes with a small $m_{q}$ for kaon in Eq.~\eqref{eq:kaonK}, 
its effect is insignificant for the interaction kernel for kaon. 
This would make $K_{K}$ reduced and the kaon channel would get more repulsive. 
To keep the kaon mass in this repulsive change, the current strange quark mass should be reduced. 

It is also interesting to mention that $G_{S}$ is less than unity. According to Ref.~\cite{Kono:2019aed},
this situation is called 
anomaly-driven symmetry breaking, where the U(1)$_{A}$ anomaly plays
an essential role for the dynamical breaking of chiral symmetry. In such a situation,
the mass of the sigma meson should be smaller than 800~MeV. 
The current model also suggests that the anomaly-driven symmetry breaking takes place and 
predicts that the sigma meson mass is found to be less than 800~MeV
as we will show in the next subsection.

\subsection{Scalar meson masses}  \label{sec:scalar}

\begin{table}
\centering
\begin{tabular}{rl|rrrrr} 
\hline\hline
$G_{k}$ & & $0.0$ & $0.5$ & $1.0$ & $1.5$ & $2.0$ \\
\hline
$\textrm{Re}(m_{a_{0}})$ & [MeV] & $ 844$ & $ 865$ & $ 883$ & $ 898$ & $ 912$ \\
$\textrm{Re}(m_{\kappa})$ & [MeV] & $ 970$ & $ 923$ & $ 886$ & $ 857$ & $ 832$ \\
$m_{\sigma}$ & [MeV] & $ 585$ & $ 584$ & $ 583$ & $ 582$ & $ 581$ \\
$\textrm{Re}(m_{\sigma^{\prime}})$ & [MeV] & $1077$ & $ 991$ & $ 925$ & $ 873$ & $ 832$ \\
$m_{\eta}$ & [MeV] & $ 533$ & $ 552$ & $ 560$ & $ 559$ & $ 554$ \\
$M_{q}$ & [MeV] & $ 309$ & $ 309$ & $ 309$ & $ 309$ & $ 309$ \\
$M_{s}$ & [MeV] & $ 475$ & $ 428$ & $ 392$ & $ 362$ & $ 338$ \\
$(-\vev{\bar qq})^{1/3}$ & [MeV] & $ 248$ & $ 248$ & $ 248$ & $ 248$ & $ 248$ \\
$(-\vev{\bar ss})^{1/3}$ & [MeV] & $ 269$ & $ 265$ & $ 261$ & $ 257$ & $ 253$ \\
$f_{K}$ & [MeV] & $  98$ & $  96$ & $  95$ & $  94$ & $  92$ \\
\hline\hline
\end{tabular}

\caption{Results of the physical quantities obtained in the present calculation. 
The cutoff is fixed as $\Lambda = 650$ MeV.} \label{tab:results}
\end{table}

Fixing the model parameters for each $G_{k}$, we calculate the meson masses for $G_{k}$ from $0.0$ to $2.0$. 
The results are summarized in Table~\ref{tab:results}.
The $\sigma$ and $\eta$ mesons are found below the threshold of the $\bar q$-$q$ pair creation 
without a decay width. This implies that they are bound states of two constituent quarks and do not decay into
the quark pair nor mesonic modes. This is because we do not consider mesonic decay channels 
in the Bethe-Salpeter equations. 
On the other hand, the $a_{0}$, $\kappa$ and $\sigma^{\prime}$
mesons are found as resonances with decay widths.
The $a_{0}$ and $\kappa$ mesons are generated in the single $\bar q$-$q$ and $\bar s$-$q$ scattering channels,
respectively, and these resonance states are located above the thresholds of these channels. 
The $\sigma^{\prime}$ mesons are in the coupled channels of 
the $\bar q$-$q$ and $\bar s$-$s$ scatterings and are found above both thresholds. 
The decay widths of these mesons are for the fall apart of the quark and antiquark
due to the lack of the confinement mechanism in the NJL model. Therefore these 
decay widths are model artifacts and we do not discuss their details. 
For this reason, in Table~\ref{tab:results} we show only the real parts of the meson masses. 
In Table~\ref{tab:results} we show also our calculated results of the constituent quark masses, $M_{q}$ and $M_{s}$,
the quark condensates, $\vev{\bar qq}$ and $\vev{\bar ss}$, and the kaon decay constant $f_{K}$.

Let us first discuss the case of $G_{k} = 0.0$, which is the original NJL model with the determinant interaction. 
The constituent quark masses, $M_{q}$ and $M_{s}$, are found as 309~MeV and 475~MeV, respectively, which are consistent with the empirical values. The kaon decay constant is underestimated in the present calculation, 
but this is known as one of the drawbacks of the NJL model \cite{Takizawa:1996nw}.
As mentioned in introduction, thanks to the heavy strange quark mass than those of the up and down quarks,
the mass of the $\kappa$ meson is heavier than that of the $a_{0}$ meson, which is the normal mass ordering 
and inconsistent with the observation. The $\sigma$ meson is reproduced with a 600 MeV mass, which is consistent 
with an empirical value. 
The $\eta$ meson mass is reproduced reasonably well, although 
it is a bit underestimated in comparison with the observed $\eta$ meson mass. 

\begin{figure}
  \centering
   \includegraphics[width=0.8\linewidth]{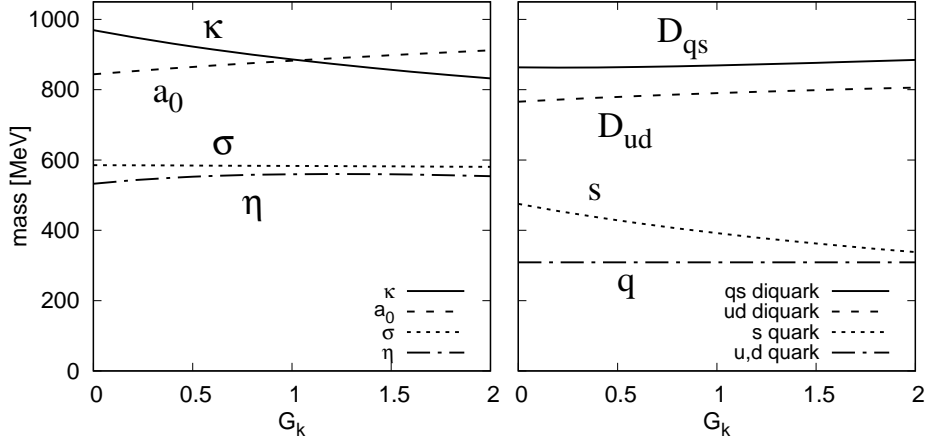}
\caption{Dependence of the masses of the mesons, the constituent quarks and the diquarks to the $G_{k}$ parameter. 
The left panel shows the meson masses, while the right plots the masses of the constituent quark 
and the diquarks.  For discussion on the diquark masses, see Sect.~\ref{sec:diquark}.
%Symbols in the key, {\arial q} and {\arial s}, stand for the constituent masses of the up-down quark and 
%the strange quark, respectively. 
The real parts of the resonance masses are plotted.  
The cutoff is fixed as $\Lambda = 650$~MeV.
} \label{fig:kterm}
\end{figure}

\begin{figure}
  \centering
   \includegraphics[width=0.8\linewidth]{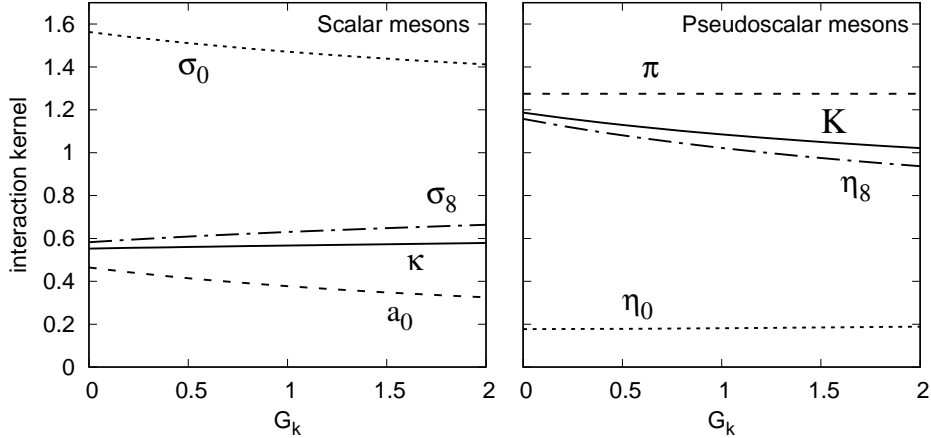}
\caption{Dependence of the determined interaction kernels in units of $g_{S}^{0}\equiv 2\pi^{2}/(3\Lambda^{2})$ 
to the $G_{k}$ parameter. 
The left panel is for the scalar meson, while the right for the pseudoscalar mesons.
The interaction kernel for each meson is given in Eq.~\eqref{KScalar} for the 
scalar mesons and Eq.~\eqref{KPScalar} for the pseudoscalar mesons. 
} \label{fig:coup}
\end{figure}

%We plot the $G_{k}$ dependence of the masses of the mesons and the constituent quarks in Fig.~\ref{fig:kterm}.
We plot the $G_{k}$ dependence of the meson masses in the left panel of Fig.~\ref{fig:kterm}.
This figure shows that, as the value of $G_{k}$ increases, the mass of the $\kappa$ meson decreases while 
the $a_{0}$ mass increases. It is remarkable that at $G_{k} \sim 1$ 
the mass of the $\kappa$ meson becomes lighter than that of the $a_{0}$ meson. Thus,  
the inverse mass order of the $\kappa$ and $a_{0}$ mesons is realized for $G_{k} > 1.0$. 
This is understood as follows. We show the $G_{k}$ dependence 
of the interaction kernels in Fig.~\ref{fig:coup}. As seen in the left panel, 
the interaction kernel of the $a_{0}$ meson decreases as $G_{k}$ increases on one hand. 
This is because, as seen in Eq.~\eqref{eq:a0K}, 
the negative contribution of the $G_{k}$ term to the interaction kernel $K_{a_{0}}$ appears 
with a large current mass of the strange quark. 
On the other hand, the interaction kernel for the $\kappa$ meson increases gradually. 
This is because, for the $\kappa$ meson, the $G_{k}$ term comes with the small current quark mass $m_{q}$
and its contribution gets minor for the interaction kernel $K_{\kappa}$. 
In this way, we confirm that the new mechanism proposed in Ref.~\cite{KHMJ} works well also in the NJL model. 

Figure~\ref{fig:kterm} shows that the $\sigma$ meson mass gradually decreases as $G_{k}$ increases.
One may wonder why the $\sigma$ meson mass decreases although  
the interaction kernel for the $\sigma$ meson gets weaker as seen in Fig.~\ref{fig:coup}. 
This is because in the $\sigma$ meson the flavor singlet component is dominated as shown in Table~\ref{tab:singoct},
and thus the strange quark is also one of the main components of the $\sigma$ meson. 
In addition, as seen in Fig.~\ref{fig:kterm},
the constituent strange quark mass gets reduced as $G_{k}$ increases, which will be discussed in the next section. 
Because the mass of the constituents gets reduced,
the mass of the $\sigma$ meson decreases even though the coupling strength gets weaker.  

\begin{table}
\centering 
\begin{tabular}{cc|rrr} 
\hline\hline
state & componet & $G_{k}=0.0$ & $G_{k}=1.0$ & $G_{k}=2.0$ \\
\hline
$\sigma$ & singlet    & $97.2\, \%$ & $96.5\, \%$ & $96.3\, \%$ \\
 & octet  & $ 2.8\, \%$ & $ 3.5\, \%$ & $ 3.7\, \%$ \\
\hline
$\sigma^\prime$ & singlet & $22.0\, \%$ & $11.3\, \%$ & $ 5.1\, \%$ \\
 & octet  & $78.0\, \%$ & $88.7\, \%$ & $94.9\, \%$ \\
\hline
$\eta^\prime$ & singlet & $98.8\, \%$ & $98.5\, \%$ & $97.5\, \%$ \\
 & octet  & $ 1.2\, \%$ & $ 1.5\, \%$ & $ 2.5\, \%$ \\
\hline
$\eta$ & singlet   & $ 0.0\, \%$ & $ 0.7\, \%$ & $ 2.1\, \%$ \\
 & octet  & $100.0\, \%$ & $99.3\, \%$ & $97.9\, \%$ \\
\hline\hline
\end{tabular}
\caption{Obtained flavor contents of the $\sigma$, $\sigma^{\prime}$, $\eta$ and $\eta^{\prime}$
mesons for $G_{k}=0.0$, $1.0$ and $2.0$. } \label{tab:singoct}
\end{table}

\subsection{Constituent quark masses} \label{sec:quark}

\begin{figure}
  \centering
   \includegraphics[width=0.6\linewidth]{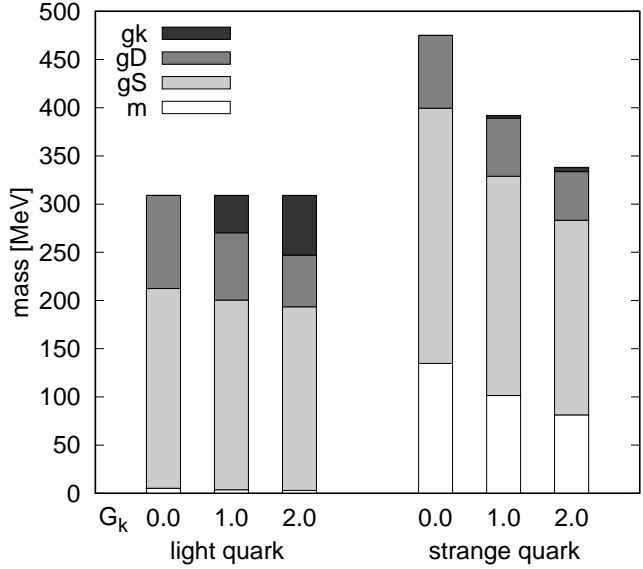}
\caption{Contents of the constituent quark masses for $G_{k}=0.0,\ 1.0,\ 2.0$. 
The constituent quark masses are decomposed accordingly to the right hand sides of the gap equations~\eqref{gapeq}. 
Symbols in the key, {\arial m}, {\arial gS}, {\arial gD} and {\arial gk}, stand for the contributions from the first, second, third and fourth terms of 
the right hand sides of Eq.~\eqref{gapeq}, 
respectively.} \label{fig:mqcont}
\end{figure}

It is also interesting noting that Fig.~\ref{fig:kterm} shows that the constituent strange quark also 
reduces its mass $M_{s}$ as the $G_{k}$ increases, while the constituent quark mass $M_{q}$ stays at the same value.
This is also one of the reasons that the $\kappa$ meson mass is reduced with the increase of $G_{k}$. 
%For the $\kappa$ meson the strange quark is one of the main ingredients. 
%If the constituent mass of the strange quark gets reduced, the $\kappa$ meson mass becomes smaller. 
In order to understand the origin of the reduction of the strange quark mass, 
let us decompose the contents of the constituent quark masses. 
The constituent quark masses are determined by the gap equations~\eqref{gapeq}. 
The quark masses can be decomposed accordingly to the right hand side of Eq.~\eqref{gapeq}, and 
we show this decomposition of the quark masses in Fig.~\ref{fig:mqcont}. 
For the light quarks, $u$ and $d$, one finds that
the contribution of the $G_{k}$ term is enhanced for lager $G_{k}$. This is because 
the $G_{k}$ term for the light quarks appears with the current strange quark mass $m_{s}$.
At the same time, the contributions from the $G_{S}$ and $G_{D}$ terms get reduced.
The contribution from the current quark mass is negligibly small for the light quarks. 
For the strange quark, the contribution of the $G_{k}$ term is negligible, 
because this term appears together with the tiny current quark mass $m_{q}$
in the gap equation for the strange quark mass. Therefore, for the strange quark, 
there is no enhanced contributions for larger $G_{k}$. In addition, the contribution
from the current quark mass gets also smaller with the reduction of the current strange quark mass.

\section{Discussion} \label{sec:discussion}

\subsection{Diquark mass}
\label{sec:diquark}

With a finite $G_{k}$ we obtain a smaller constituent strange quark mass. 
For $G_{k} \simeq 2.0$ where the inverse mass ordering is realized, 
the constituent strange quark gets almost degenerate to the constituent up-down quark.
This means that the SU(3) breaking on the constituent quark masses gets weaker.
In such a situation, one may wonder whether this degeneracy of the constituent quarks
would be inconsistent with the baryon mass spectra, where substantial SU(3) flavor 
breaking is present. Here, instead of calculating the baryon masses directly in
Faddeev approaches developed in Refs.~\cite{Buck:1992wz,Ishii:1993np,Ishii:1993rt,Ishii:1995bu,Huang:1993yd},
rather we calculate the diquark masses, because the diquarks are significant ingredients of the baryons 
and the baryon masses can be expressed in terms of the diquark mass together with the constituent quark mass~\cite{Vogl:1990fh,Suzuki:1992pa,Weiss:1993kv}.

\begin{table}
\centering
\begin{tabular}{c}
\hline\hline
%\multicolumn{2}{c}{Scalar diquarks}\\
Scalar diquarks\\
\hline
$\Phi_{ud}^{a} = \frac{\epsilon_{abc}}{\sqrt 2}( u_{b}^{T}C \gamma_{5} d_{c} - d_{b}^{T}C \gamma_{5} u_{c})$, \\ 
$\Phi_{qs}^{a} = \frac{\epsilon_{abc}}{\sqrt 2}( q_{b}^{T}C \gamma_{5} s_{c} - q_{b}^{T}C \gamma_{5} s_{c})$, \\
\hline\hline
\end{tabular}
\caption{Quark contents of the diquark interpolating fields. Indices $a,b,c$ are for 
the color space, $q$ is the isodoublet quark field and $C$ is the charge conjugation matrix.}  
\label{tab:diqQC} 
\end{table}

We consider flavor-antisymmetric scalar diquarks, so-called good diquarks, 
which are the most attractive channels among the diquarks. The interpolating fields for the scalar diquarks 
are shown in Table~\ref{tab:diqQC}. 
To obtain the scalar diquark masses,
we calculate the pole position of the scattering matrix of the two quark system using the Bethe-Salpeter
equation~\eqref{BSeq}. 
The interaction kernels for the diquark channels are also obtained from Lagrangian~\eqref{NJLLag}
by performing a Fierz transformation. But it is known that the Fierz transform of the $g_{S}$ term 
in Lagrangian~\eqref{NJLLag} does not provides the scalar diquark interactions.
Here we assume that the strength of the $g_{S}$ term for the scalar diquark channels is 
a quarter of that of the pseudoscalar meson channel, which is obtained in the Fierz transformation 
of the color current interaction ${\cal L}_\textrm{int} = - G_{c} \sum_{a=1}^{8} (\bar \psi \gamma_{\mu} t^{a} \psi)^{2}$,
where $t^{a}$ is the generator of color space. 
It is our aim to see the dependence of the diquark mass to the $G_{k}$ parameter, while 
we are not interested in the quantitative reproduction of the diquark masses. Thus,
the exact value of the coefficient of the $g_{S}$ term is irrelevant to our discussion.  
The $g_{D}$ and $g_{k}$ terms for the diquark channels can be obtained by the Fierz transformation 
of Lagrangian~\eqref{NJLLag}, and then we have the interaction kernel of the diquark channels as 
\begin{subequations} \label{Kdiquark}
\begin{align}
K_{ud} &= \frac14 g_{S} + \frac 12 g_{D}\vev{\bar ss} + \frac14 g_{k} m_{s},  \label{eq:udK} \\ 
K_{qs} &= \frac14 g_{S} + \frac 12 g_{D}\vev{\bar qq} + \frac14 g_{k} m_{q}. \label{eq:qsK}
\end{align}
\end{subequations}

The loop function for the diquark channel is defined by the correlation of the interpolating fields $\Phi_{D}^{a}(x)$
given in Table~\ref{tab:diqQC} with (anti)color $a$ and is obtained as the same form as the loop function 
for the pseudoscalar meson channel:
\begin{equation}
   J_{D}^{ab}(W) = i \int d^{4}x\, e^{i P\cdot x}\langle 0 | T [\Phi_{D}^{a}(x) \Phi_{D}^{b}(0)] | 0 \rangle
   = \delta^{ab}\, \frac{2}{3} J_{P}(W),
\end{equation}
with $P^{\mu}=(W,0,0,0)$. The loop functions for the $ud$ and $qs$ diquarks are obtained as
\begin{subequations}  \label{Jdiquark}
\begin{align}
J_{ud}(W) &= \frac23 J_\textrm{P}(W; M_{q}, M_{q}),\\  
J_{qs}(W)  &= \frac23 J_\textrm{P}(W; M_{q}, M_{s}),
\end{align}
\end{subequations}
respectively. 

The diquark mass is obtained as a pole position of the $T$ matrix by solving Eq.~\eqref{eq:mass}.
The calculated masses of the $ud$ and $qs$ scalar diquarks are plotted in the right panel of Fig.~\ref{fig:kterm}.
Here we are interested in the flavor SU(3) breaking in the diquark masses, namely the mass difference 
between the $ud$ and $qs$ diquarks. The adjustment of their absolute values is out of the current scope.
If one would need to reproduce diquark masses, one could tune the coefficient of the $g_{S}$ term, which 
acts on the interaction kernels in a flavor symmetric way. The diquark masses are found above 
the $qq$ and $qs$ thresholds, and thus, the diquarks have a decay width. In the figure we show their real parts. 
We find %in the right panel of Fig.~\ref{fig:kterm}
that the mass difference of the diquark masses is insensitive to the $G_{k}$ parameter. 
This means that the flavor SU(3) breaking in the diquark mass is still substantially large even for a finite $G_{k}$.
Therefore, this suggest that the baryon masses could have a substantial flavor SU(3) breaking
even if the constituent quark masses get degenerate.

\subsection{Gell-Mann Oakes Renner relation}
By introducing the anomaly induced interaction with the current quark mass into the model,
the symmetry breaking pattern in this model is somewhat different from that in the original theory. 
The divergence of the axial vector current  $A_{\mu}^{a} = \bar \psi \gamma_{\mu} \gamma_{5} \frac{\lambda^{a}}{2} \psi$
can be calculated by the Noether theorem:
\begin{equation}
   \partial^{\mu}A_{\mu}^{a} = -i [  {\cal L}_\textrm{NJL}, Q_{A}^{a}],
\end{equation}
where $Q_{A}^{a}$ is the generator of the axial transformation labeled by $a$. 
In the current model, the divergence of the third component of the axial vector current $A_{\mu}^{3}$
is calculated as 
\begin{equation}
    \partial^{\mu}A_{\mu}^{3} = m_{q} P^{3} - g_{k} m_{q} \left( P^{3}\, \bar ss + S^{3}\, \bar s i\gamma_{5}s
    - \bar s i \gamma_{5} u\, \bar u s - \bar u i\gamma_{5} s \, \bar su
    + \bar s i \gamma_{5} d\, \bar d s + \bar d i\gamma_{5} d \, \bar sd \right), \label{eq:PCAC}
\end{equation}
where the pseudoscalar and scalar fields, $P^{a}$ and $S^{a}$, 
are defined as $P^{a} = \bar \psi i \gamma_{5} \lambda^{a} \psi$ and $S^{a} = \bar \psi \lambda^{a} \psi$,
respectively. This is the PCAC relation in this model and is different from that of QCD due to the 
presence of the $g_{k}$ term.

The Gell-Mann Oakes Renner (GOR) relation~\cite{Gell-Mann:1968rz} is derived by combining the Glashow Weinberg
relation~\cite{Glashow:1967rx} and the PCAC relation. (See, for instance, Ref.~\cite{Jido:2008bk}.)
The Glashow Weinberg relation is a relation in the chiral limit and reads for the massless pion 
\begin{equation}
   F_{\pi} G_{\pi} = - 2 \vev{\bar qq}, \label{eq:GW}
\end{equation}
with the matrix elements defined by
\begin{align}
   \vev {0 | A_{\mu}^{a}(x) | \pi^{b} (p)} & = i \delta^{ab} p_{\mu} F_{\pi} e^{-ip\cdot x}, \label{eq:axial} \\
   \vev {0 | P^{a}(x) | \pi^{b} (p)} & =  \delta^{ab} G_{\pi} e^{-ip\cdot x}.
\end{align}
Taking the derivative of Eq.~\eqref{eq:axial}, we obtain, on one hand,  
\begin{equation}
   \vev{0 | \partial^{\mu} A_{\mu}^{a}(x) | \pi^{b} (p)}  =  \delta^{ab} m_{\pi}^{2} F_{\pi} e^{-ip\cdot x},
\end{equation}
where the on-shell condition $p^{2} = m_{\pi}^{2}$ is taken for pion. On the other hand,
taking the matrix element for the vacuum and pion of the PCAC relation~\eqref{eq:PCAC}, 
we obtain
\begin{equation}
   \vev{0 | \partial^{\mu} A_{\mu}^{3}(x) | \pi^{0} (p)} = m_{q} G_{\pi} ( 1 - g_{k} \vev{\bar ss}) e^{-ip\cdot x},  
   %= \delta^{ab} m_{\pi}^{2} F_{\pi} e^{-ip\cdot x}
\end{equation}
for $\pi^{0}$. Thus, from the PCAC relation we obtain  
\begin{equation}
   m_{\pi}^{2} F_{\pi} = m_{q} G_{\pi} ( 1 - g_{k} \vev{\bar ss}). \label{eq:fromPCAC}
\end{equation}
Eliminating $G_{\pi}$ from Eqs.~\eqref{eq:GW} and \eqref{eq:fromPCAC}, we obtain
the Gell-Mann Oakes Renner relation for this model as
\begin{equation}
   m_{\pi}^{2} F_{\pi}^{2} = - 2 m_{q} \vev{\bar qq} (1 - g_{k} \vev{\bar ss}). \label{eq:GOR}
\end{equation}
Taking $g_{k}=0$, one sees that the original form of the GOR relation is recovered. 
In the numerical calculations given in Sect.~\ref{sec:results}, 
the GOR relation~\eqref{eq:GOR} is satisfied with more than 99\% precision, which 
is estimated by the ratio of the left hand side to the right hand side of Eq.~\eqref{eq:GOR} with 
the values obtained in our calculation. 

Actually, Eq.~\eqref{eq:GOR} is consistent with the GOR relation in QCD. 
The current model is an effective model of QCD at low energy. 
According to the idea of the effective field theory, one presumes that the partition functions $Z[v,a,s,p]$
defined through the path integral form with external fields $s,a,s,p$ should have common symmetry properties 
and should be shared by the original theory
and the effective theory at low energy:
\begin{equation}
  Z[v,a,s,p] = \int {\cal D} q {\cal D}\bar q {\cal D} G_{\mu} e^{i \int d^{4}x {\cal L}_\textrm{QCD}[v,a,s,p]}
  = \int {\cal D} \psi {\cal D}\bar \psi e^{i \int d^{4}x {\cal L}_\textrm{NJL}[v,a,s,p]}.
\end{equation}
In this formulation, the quark condensate is obtained by
\begin{equation}
   \vev{\Omega | \bar qq | \Omega } = -i \frac{1}{Z[0]} \left. \frac{\delta Z}{\delta s} \right|_{s=m_{q}, v=a=p=0}.
\end{equation}
Calculating the above functional derivative for both theories in the mean field approximation, 
we obtain
\begin{equation}
   \vev{ \bar qq}_\textrm{QCD} = \vev{\bar qq} ( 1- g_{k} \vev{\bar ss}), \label{eq:qbarq}
\end{equation}
where the left hand side denotes the quark condensate for QCD. This means that the mean fields 
in the current model may different from those in QCD. Combining Eqs.~\eqref{eq:GOR} and \eqref{eq:qbarq},
one restores the original GOR relation for QCD. 
{Equation~\eqref{eq:qbarq} corresponds to Eq.~(15) of Ref.~\cite{KHMJ} 
that was obtained in the linear sigma model.}

%%%%%%%%%%%%

\section{Summary} \label{sec:summary}

We have investigated the scalar meson masses in the Nambu Jona-Lasinio model 
by solving the Bethe Salpeter equation for the quark-antiquark channels. 
Motivated by the previous work for the scalar meson masses in a linear $\sigma$ model~\cite{KHMJ},
we have introduced an axial anomaly induced interaction containing the current quark mass.
Without this term, 
{the $\bar qq$ scalar meson spectrum has an inconsistent ordering 
with observation that the $\kappa$ meson with strangeness 
were heavier than the $a_{0}$ meson with isospin $I=1$.  }
The anomaly induced interaction with the current quark mass gives repulsion to the quark and antiquark 
systems for the flavor octet
scalar channels, and it comes to the $a_{0}$ meson with a sizable strange current quark mass, 
while it does to the $\kappa$ meson with a tiny up or down current quark mass. 
This different influence induces the reversal of the mass order between the $a_{0}$ and $\kappa$ mesons.  
By fixing the model parameters to reproduce the masses of pion, kaon and the $\eta^{\prime}$ meson 
and the pion decay constant, we have calculated
the scalar meson masses. We have found that 
around $G_{k} \simeq 1.0$ the masses of the $a_{0}$ and $\kappa$ mesons get degenerate 
and for $G_{k} > 1.0$ the $a_{0}$ meson mass gets heavier than that of the $\kappa$ meson.
Thus, the inverse mass ordering is realized even in the $\bar qq$ configuration for the scalar meson. 
It is also interesting to mention that this model suggests that the constituent quark masses get degenerate
for $G_{k} \simeq 2.0$.  This might be inconsistent with the baryon mass spectra in which one sees 
substantial flavor SU(3) breaking. By calculating the scalar diquark masses in the current model, 
we have found that the scalar diquark masses have still substantial flavor breaking even if the 
constituent quark masses get degenerate. Therefore, if the baryons have considerably large amount 
of the diquark component in their structure, the baryon masses still have enough flavor breaking 
even if the constituent quark masses get degenerate. 

In conclusion, by introducing the U(1) axial anomaly term with the current quark mass,
it is possible to realize the inverse mass ordering of the light scalar meson even 
in the $\bar qq$ picture. This fact would suggest that 
the $f_{0}(500)$ meson could be the chiral partner of the pion or at least should have 
a substantial component of the chiral partner. 

\section*{Acknowledgements}
%\acknowledgments
The authors would like to thank Prof.\ Fujioka for his important suggestions. 
The work of D.J.\ was partly supported by Grants-in-Aid for Scientific Research from JSPS 21K03530.
The work of M.H. is supported in part by 
JPSP KAKENHI Grant Number 20K03927.
%%%

\appendix
\section{Calculation details}

In this appendix we show the calculation details of the formulae used in the NJL model. 
We perform the loop integrals with the three momentum cutoff $\Lambda$.

The quark condensate is calculated as
\begin{align}
%\lefteqn{
   \left\langle \bar{\psi} \psi \right\rangle &
   %= - i N_c {\lim_{x \to 0}} \textrm{ Tr} [S_{F}(x)]   %&\nonumber \\&
%\nonumber \\&
   = -  \frac{N_c M}{\pi^{2}} \int_{0}^{\Lambda} \frac{p^{2}}{\sqrt{p^{2} + M^{2}}}dp
   %}
%   \nonumber \\&
    = - \frac{N_c M\Lambda^{2}}{2\pi^{2}} 
  \left(\sqrt{1+\frac{M^{2}}{\Lambda^{2}}} 
  - \frac{M^{2}}{\Lambda^{2}} \log \frac{\Lambda + \sqrt{ \Lambda^{2} + M^{2}}}{M} \right).
  \label{eq:SF}
\end{align}
%with $ x = M/\Lambda$. 

The loop functions for the scalar and pseudoscalar channels in the center of mass frame
$P^{\mu} = (W, 0,0,0)$ are show for the energy below the threshold, 
$W \le M_{1} + M_{2}$, as
\begin{align}
  J_\textrm{S}(W;M_{1},M_{2}) 
   =& N_{c} i \int \frac{d^{4}p}{(2\pi)^{4}} \textrm{Tr} 
   \left[\frac{1}{\diracslash{p} - M_{1}+i\epsilon}\frac{1}{\diracslash{p} - \Pslash - M_{2}+i\epsilon}\right] \nonumber \\
   =&  \frac{N_{c}\Lambda^{2}}{4\pi^{2}} \left\{
    \sqrt{1 + \frac{M_{1}^{2}}{\Lambda^{2}}} 
    - \frac{M_{1}^{2}+M_{2}^{2}-W^{2}}{2\Lambda^{2}} \log\frac{\Lambda + \sqrt{\Lambda^{2} + M_{1}^{2}}}{M_{1}}
  \right.  \nonumber \\ 
  & -  \frac{(M_{1}+M_{2})^{2}}{\Lambda^{2}} 
   \frac{\omega_{1}}{W} \log\frac{\Lambda + \sqrt{\Lambda^{2} + M_{1}^{2}}}{M_{1}}
   \nonumber \\  & \left.
   -\frac{W^{2}-(M_{1}+M_{2})^{2}}{\Lambda^{2}} \frac \eta {W} \tan^{-1} \frac{\omega_{1}}{\eta} \sqrt{\frac{\Lambda^{2}}{\Lambda^{2}+M_{1}^{2}}} \right\} + (M_{1} \leftrightarrow M_{2}),  \label{eq:JS} \\
%\lefteqn{J_{\textrm{PS},K}(\sqrt s;M_{1},M_{2})} \nonumber \\
J_{\textrm{P}}(W;M_{1},M_{2})
=& N_{c} i \int \frac{d^{4}p}{(2\pi)^{4}} 
  \textrm{Tr}\left[i\gamma_{5}\frac{1}{\diracslash{p} - M_{1}+i\epsilon} i\gamma_{5} \frac{1}{\diracslash{p} - \Pslash - M_{2}+i\epsilon}\right] \nonumber \\
%=& \frac{N_{c} \Lambda^{2}}{4\pi^{2}} 
%  \left\{ \sqrt{1+x_{1}^{2}} - \frac{x_{1}^{2} + x_{2}^{2} -y^{2}}{2} \log \frac{1+\sqrt{1+x_{1}^{2}}}{x_{1}}
%  \right. \nonumber \\
%  &- (x_{1}-x_{2})^{2} \frac{\omega_{1}}{y} \log \frac{1+\sqrt{1+x_{1}^{2}}}{x_{1}}
%  \nonumber \\ 
%  & \left. - \left[ y^{2} - (x_{1} - x_{2})^{2} \right]
%   \frac{\eta}{y} \tan^{-1} \frac{\omega_{1}}{\eta \sqrt{1+x_{1}^{2}}} \right\}
%  \nonumber \\
%  & + (x_{1} \leftrightarrow x_{2}) 
=& \frac{N_{c} \Lambda^{2}}{4\pi^{2}} 
  \left\{ \sqrt{1+\frac{M_{1}^{2}}{\Lambda^{2}}} - \frac{M_{1}^{2} + M_{2}^{2} -W^{2}}{2\Lambda^{2}} 
  \log \frac{\Lambda+\sqrt{\Lambda^{2}+M_{1}^{2}}}{M_{1}}
  \right. \nonumber \\
  &- \frac{(M_{1}-M_{2})^{2}}{\Lambda^{2}} \frac{\omega_{1}}{W} \log \frac{\Lambda+\sqrt{\Lambda^{2}+M_{1}^{2}}}{M_{1}}
  \nonumber \\ 
  & \left. - \frac{ W^{2} - (M_{1} - M_{2})^{2}}{\Lambda^{2}}
   \frac{\eta}{W} \tan^{-1} \frac{\omega_{1}}{\eta} \sqrt{\frac{\Lambda^{2}}{\Lambda^{2}+M_{1}^{2}}} \right\}
  %\nonumber \\ 
  + (M_{1} \leftrightarrow M_{2}),
 \label{eq:JP}
\end{align}
where we have defined %$x_{1} = M_{1}/\Lambda$, $x_{2} = M_{2}/\Lambda$, 
\begin{align}
% & \omega_{1} = \frac{y^{2} + x_{1}^{2} - x_{2}^{2}}{2y}, \quad
%  \omega_{2} = \frac{y^{2} + x_{2}^{2} - x_{1}^{2}}{2y}, \\
% & \eta = \frac{\sqrt{[(x_{1}+x_{2})^{2} - y^{2}][y^{2}- (x_{1}-x_{2})^{2}]}}{2y}.
 & \omega_{1} = \frac{W^{2} + M_{1}^{2} - M_{2}^{2}}{2W}, \quad
  \omega_{2} = \frac{W^{2} + M_{2}^{2} - M_{1}^{2}}{2W}, \\
 & \eta = \frac{\sqrt{[(M_{1}+M_{2})^{2} - W^{2}][W^{2}- (M_{1}-M_{2})^{2}]}}{2\sqrt s}.
\end{align}
The analytic continuation from $W < M_{1}+ M_{2}$ to $W > M_{1}+M_{2}$ can be
performed as 
\begin{align}
   \frac{\eta}{W} \tan^{-1} \frac{\omega_{1}}{\eta \sqrt{1 + M_{1}^{2}/\Lambda^{2}}}
   =
   \frac{q}{2W} \log \frac{ q \sqrt{1 + \frac{M_{1}^{2}}{\Lambda^{2}}} + \omega_{1}}{q \sqrt{1+\frac{M_{1}^{2}}{\Lambda^{2}}} - \omega_{1}} ,
\end{align}
with 
\begin{equation}
  q = \frac{\sqrt{(W^{2}- (M_{1}+M_{2})^{2})(W^{2}-(M_{1}-M_{2})^{2})}}{2\sqrt s}.
\end{equation}

The loop function with the axial current is calculated 
%\begin{align}
%    P^{\mu} J_{A,K}(P) & \equiv i N_{c} \int \frac{d^{4}p}{(2\pi)^{4}} 
%   \textrm{Tr}\left[\frac{\gamma^{\mu}\gamma_{5}}2 \frac{1}{\diracslash{p}-M_{1}} 
%   i \gamma_{5} \frac{1}{\diracslash{p} - \Pslash- M_{2}}\right],
%\end{align}
%and at $P=(\sqrt s, 0,0,0)$
\begin{align}
   %\lefteqn{
   J_{A}(W; M_{1}, M_{2}) =&\ 
    N_{c} i\int \frac{d^{4}p}{(2\pi)^{4}} 
   \textrm{Tr}\left[\frac{\gamma^{\mu}\gamma_{5}}2 \frac{1}{\diracslash{p}-M_{1}+i\epsilon} 
   i \gamma_{5} \frac{1}{\diracslash{p} - \Pslash- M_{2} + i\epsilon }\right] \nonumber \\
%   & \nonumber \\& \quad 
   =&\ -\frac{iN_{c}\Lambda}{4\pi^{2}}
  \left\{
  \frac{\Lambda(M_{2}-M_{1})}{2 W^{2}} \sqrt{1 + \frac{M_{1}^{2}}{\Lambda^{2}}} 
%   \nonumber \right. \\ & \left. \qquad
   +\frac{M_{1}}\Lambda \frac{\omega_{1}}{W} 
      \log \frac{\Lambda + \sqrt{\Lambda^{2} + M_{1}^{2}}}{M_{1}}
    \right. \nonumber \\ & \left. \quad
   + \frac{M_{1}-M_{2}}\Lambda \frac{\eta^{2}-\omega_{1}^{2}}{2W^{2}}  
      \log \frac{\Lambda + \sqrt{\Lambda^{2} + M_{1}^{2}}}{M_{1}}
   \nonumber \right. \\ & \left. \quad
     + \frac{M_{1}+M_{2}}\Lambda \frac{(M_{1}-M_{2})^{2}-W^{2}}{W^{2}} \frac{\eta}{2W} \tan^{-1} \frac{\omega_{1}}{\eta} \sqrt{\frac{\Lambda^{2}}{\Lambda^{2}+M_{1}^{2}}}\right\}
     \nonumber \\ &  
     +  (M_{1} \leftrightarrow M_{2}).  \label{eq:JA}
\end{align}

%%%%%%%%%%%%%%%%%%

%% The Appendices part is started with the command \appendix;
%% appendix sections are then done as normal sections
%% \appendix

%% \section{}
%% \label{}

%% If you have bibdatabase file and want bibtex to generate the
%% bibitems, please use
%%
%%  \bibliographystyle{elsarticle-num} 
%%  \bibliography{<your bibdatabase>}

%% else use the following coding to input the bibitems directly in the
%% TeX file.

\end{document}